\documentclass[letterpaper, 10 pt, conference]{ieeeconf}  

\IEEEoverridecommandlockouts                              

\usepackage{cite}
\usepackage[utf8]{inputenc}
\usepackage[utf8]{inputenc}
\usepackage[english]{babel}
 \usepackage[utf8]{inputenc} 
\usepackage[T1]{fontenc}    
\usepackage{hyperref}       
\usepackage{url}            
\usepackage{booktabs}       
\usepackage{amsfonts}       
\usepackage{nicefrac}       
\usepackage{microtype}      
\usepackage{lipsum}
\usepackage{amsmath}
\usepackage[utf8]{inputenc}
\usepackage[english]{babel}
\usepackage{graphicx}
\usepackage{amssymb}
\usepackage{bigints}
\usepackage{graphicx}
\usepackage{epstopdf}
\usepackage{amssymb}
\usepackage{caption}
\usepackage{subcaption}
\usepackage{stmaryrd}
\usepackage[dvipsnames]{xcolor}
\usepackage{lipsum}
\usepackage{mathtools}
\usepackage{cuted}

\title{\LARGE \bf
Dynamic Instability of Follower Forced Euler Bernoulli Cantilever Beam With Tip Mass

}

\author{Premjit Saha$^{1}$ and Tarunraj Singh$^{2}$
\thanks{$^{1}$, PhD. Mechanical and Aerospace Engineering,
        University at Buffalo, NY, USA
        {email:~\tt\small premjits@buffalo.edu}}%
\thanks{$^{2}$Professor Mechanical and Aerospace Engineering,
        University at Buffalo, NY, USA
        {email:~\tt\small tsingh@buffalo.edu}}%
}

\begin{document}

\maketitle
\thispagestyle{empty}
\pagestyle{empty}

\begin{abstract}

This work focuses on the stability analysis of an Euler Bernoulli cantilever beam with a tip mass at the free end, subject to a follower force. This can serve as a viable model for analysis of elastic instability occurring due to fluid-structure interaction of structural components submerged in fluids and gases. A linear model with appropriate boundary conditions is developed using the energy formulation. The characteristic equation of the linear model establishes the relationship between the pulsation of the beam and the magnitude of applied follower force. The evolution of temporal eigenvalues with respect to the magnitude of the follower force helps in evaluation of the critical follower forces responsible for different modes of instability. The presented model demonstrates the existence of only dynamic instability in the system. Furthermore, the model predicts that both types of the dynamic instability i.e., flutter and divergence, are possible in the system.  


\end{abstract}

\section{INTRODUCTION}
Elastic instability of a cantilevered beams due to a non-conservative follower force can be a useful model to predict instabilities demonstrated in systems such as aircraft wings subject to the engine thrust, rockets, and in garden hoses. A vast amount of work has been done since Beck \cite{Beck.1952}, Pflüger \cite{Pfluger.1955, Pfluger.1964}, Ziegler \cite{Ziegler.1952, Ziegler.1977}, and Bolotin \cite{Bolotin.1965, J.1964}  formulated the elastic stability problems of structures due to follower force. There has been a debate on the feasibility and existence of follower forces, Yoshihiko et al. \cite{Sugiyama1999RealisticFF} and Elishakoff \cite{10.1115/1.1849170} have advocated for the realistic feasibility and existence of follower force. Paidoussis \cite{paidoussis1998fluid} presented a systematic Galerkin based method to formulate the elastic stability problem under follower force as an eigenvalue problem, where mode shapes from free vibration problem are considered as basis function. Bigoni et al.~\cite{Bigoni.2018} design and fabricate a ingenious setup using a treadmill to generate and demonstrate flutter and divergence instabilities in a Pflüger column. In the present work, the eigen-frequency and mode shape are derived from the exact characteristic equation. Majority of the existing works deal with the instability of beam under follower force either as an eigenvalue problem or as an exercise to evaluate critical load from characteristic equation considering only the first few modes. In this present work, both approaches are used to evaluate critical load for instability and a large number of modes have been considered for accuracy.

\section{Mathematical formulation of the problem}

\begin{figure}[htp]
\centering
{\includegraphics[width=0.4825\textwidth]{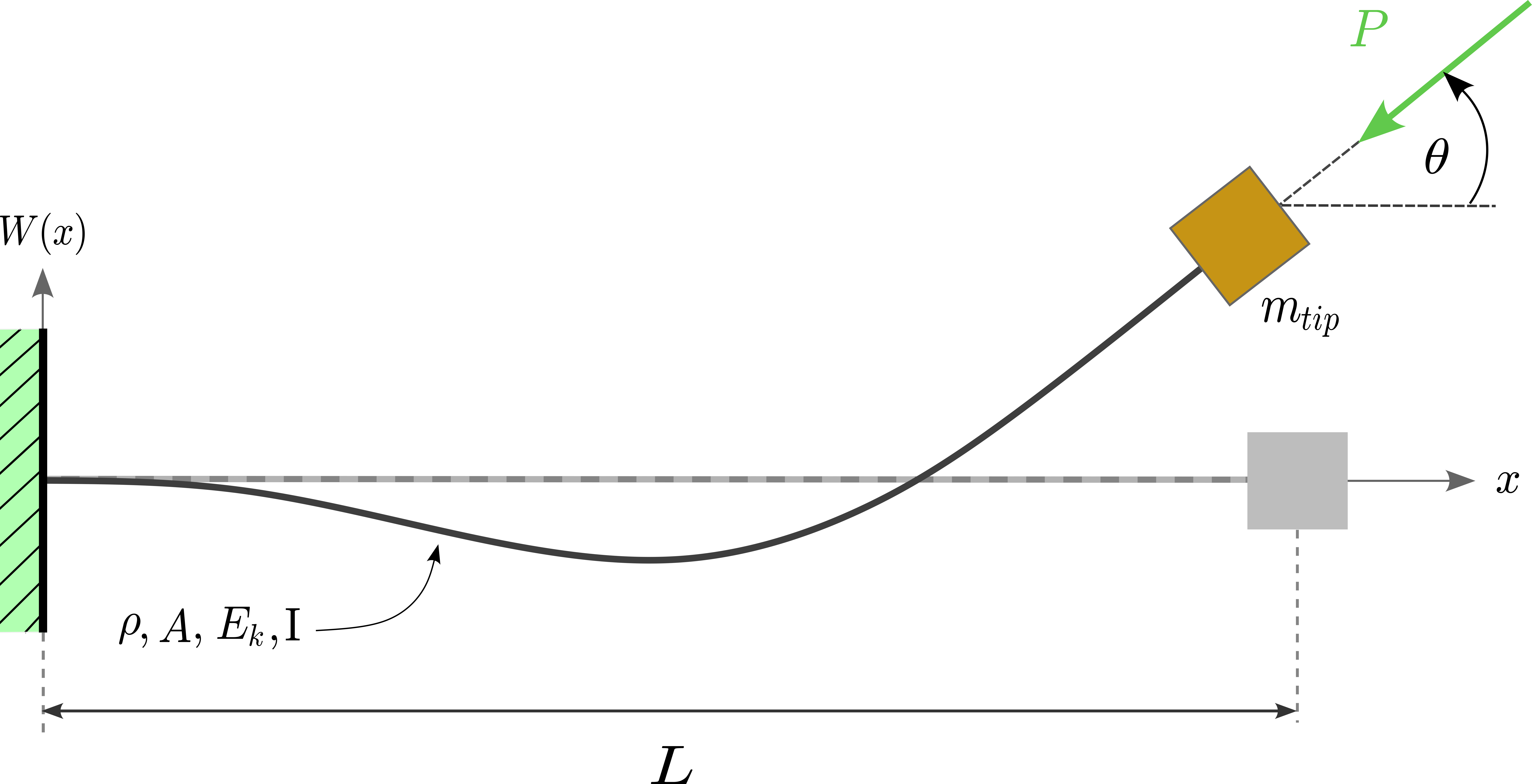}}%
\caption{Schematics of Euler Bernoulli beam under follower force }
\label{schematics_Euler_Bernoulli}
\end{figure}

In this section, a formulation for vibration of an Euler-Bernoulli beam of length $L$ under follower force will be presented in detail. Figure~\ref{schematics_Euler_Bernoulli} illustrates a clamped free beam with a load $P$ applied at the free end with the direction of the force being coincident with the slope of the end of the beam. There is also a tip mass $m_{tip}$ at the free end of the beam. Now in this particular problem, it is assumed that Young's modulus ($E_k$), density ($\rho$), and geometric properties ($I, A$) of the beam are invariant both throughout the beam length and for all time.

\subsection{Governing equation and boundary conditions}

Beck \cite{Beck.1952} first derived the governing equation for a tangentially loaded cantilevered rod. Mierovich \cite{meirovitch1967analytical} and Inman \cite{inman2001engineering} have provided both Dirichlet and Neumann boundary conditions for a beam with clamped-free boundaries with a tip mass. The governing equation for the present problem is:

\begin{equation}
    \rho A ~W_{tt}~+~E_k I ~W_{x x x x} ~+~P~W_{xx} ~=~ 0 \label{eqn9}
\end{equation}

with the boundary conditions

\begin{subequations}
\begin{equation}
    W(x,t)|_{x=0} ~=~ 0 \label{eqn12a} \\
\end{equation}
\begin{equation}
    W_x (x,t)|_{x=0} ~=~ 0 \label{eqn12b} \\
\end{equation}
\begin{equation}
    -~E_k I ~W_{x x}\Big |_{x=L}~=~0 \label{eqn12c} \\
\end{equation}
\begin{equation}
   E_k I~W_{xxx} (x,t) \Big |_{x=L}~=~m_{tip} W_{tt}(x,t) \Big |_{x=L}\label{eqn12d} 
\end{equation}
\end{subequations}     

\noindent
Here $W(x,t)$ represents the lateral deflection of the beam and is a function of the spatial location $x$ and time $t$, while $(~)_t, (~)_{tt}, ...$ and $(~)_x, (~)_{xx}, ...$ represent $\frac{\partial (~)}{\partial t} ~ , ~ \frac{\partial^2 (~)}{\partial t^2} , ... $ and $\frac{\partial (~)}{\partial x} ~ , ~ \frac{\partial^2 (~)}{\partial x^2} , ... $, respectively.

\vspace{0.125 in}
Since the governing partial differential equations (PDEs) and boundary conditions (BCs) are established (Eq.(\ref{eqn9}) and Eq.(\ref{eqn12a}-\ref{eqn12d}) respectively), a general form of solution of the governing PDE by method of separation of variables is:

       \begin{equation}
         W(x,t) ~=~\phi (x)~q (t),
       \label{eqn16}\end{equation}
       
 which transforms the PDE given by Eq.~(\ref{eqn9}) to:
       
       \begin{equation}
           \begin{split}
               & E_k I~\phi '''' (x)~q (t) + \rho A ~\phi(x)~\ddot{q} (t) \\
               & + P~\phi '' (\hat{x})~q (\hat{x}) = 0 \\
               \\
               & \Rightarrow \displaystyle \frac{E_k I \phi '''' (\hat{x})
               + P \phi '' (x)}{\rho A~\phi (x)} = - \frac{\ddot{q} (t)}{q (t)} = \beta^4 = \omega^2
           \end{split} \label{eqn17}
       \end{equation}      

Eq.(\ref{eqn17}) yields two independent ODEs
        
        \begin{equation}
            E_k I~\phi '''' (x)~+~P~\phi '' (x)~-~\beta^4 \rho A~\phi (x)=~0
       \label{eqn18} \end{equation}
        
        \begin{equation}
            \ddot{q} (t)~+~\omega^2~q (t)~=~0
        \label{eqn19} \end{equation}

\vspace{0.125 in}        
\noindent
Here $\dot{(~)}, \ddot{(~)}, ...$ and $(~)', (~)^{''}, ...(~)^{''''}$ represent $\frac{d (~)}{d t} ~ , ~ \frac{d^2 (~)}{d t^2} , ... $ and $\frac{d (~)}{d x} ~ , ~ \frac{d^2 (~)}{d x^2} , ... \frac{d^4 (~)}{d x^4}$ respectively.

\vspace{0.125 in}
\noindent
The general solution for Eq.(\ref{eqn18}) is:
      \begin{equation}
        \begin{split}
              \phi (x)~=~& B_1 ~\cosh (\lambda_1~x)~+~B_2~\sinh (\lambda_1~x) \\
                & ~+~  B_3~\cos (\lambda_2~x)~+~B_4~\sin (\lambda_2~x)  \\
                \\
           \text{where,} \quad &  \lambda_1~=~\sqrt{~\frac{~\sqrt{P^2~+~4 E_k I \rho A~\beta^4}~-~P}{2 E_k I}}~,\quad  \\ 
            & \lambda_2~=~\sqrt{~\frac{~\sqrt{P^2~+~4 E_k I \rho A ~\beta^4}~+~P}{2 E_k I}} 
        \label{eqn22}\end{split}
      \end{equation}

\vspace{0.125 in}
\noindent
Here, $\omega \in \mathbb{R}$ is eigen-frequency/the frequency of pulsation for a respective mode defined by the eigen-function $\phi(x)$. Since every eigen-function must satisfy all the boundary  conditions Eq.(\ref{eqn12a}-\ref{eqn12d}), the following set of algebraic constraints are derived after substituting $\phi~(x)$ in each boundary condition and some minor simplification post substitution.

\begin{subequations}
\begin{equation}
    B_1~+~B_3~=~0 \label{eqn23a}
\end{equation}
\begin{equation}
   \lambda_1~ B_2~+~\lambda_2~B_4~=~0 \label{eqn23b}
\end{equation}
\begin{equation}
\begin{split}
    &     \lambda_1^2~(~B_1~\cosh{(\lambda_1 L)}~+~B_2~\sinh{(\lambda_1 L)}~)~ \\
    & -~\lambda_2^2~(~B_3~\cos{(\lambda_2 L)}~+~B_4~\sin{(\lambda_2 L)}~)=~0 
\end{split}\label{eqn23c}
\end{equation}
\begin{equation}
   \begin{split}
       &  E_k I\Big(\lambda_1^3~(~B_1~\sinh{(\lambda_1 L)}~+~B_2~\cosh{(\lambda_1 L)}~)~ \\
       & +~\lambda_2^3~(~B_3~\sin{(\lambda_2 L)}~-~B_4~\cos{(\lambda_2 L)}~) \Big) \\
       & +~\beta^4 m_{tip}~\Big( B_1~\cosh{(\lambda_1 L)}~+~B_2~\sinh{(\lambda_1 L)} \\
       &~+~B_3~\cos{(\lambda_2 L)}~+~B_4~\sin{(\lambda_2 L)}~\Big)~=~0
   \end{split} \label{eqn23d}
\end{equation}
\end{subequations}

\vspace{0.125 in}
\noindent
The scalar equations, Eq.(\ref{eqn23a}-\ref{eqn23d}) can also be written in matrix vector form given by Eq.~\ref{eqn24}.

\vspace{0.95 in}

\begin{strip}
    \begin{align}
   \textstyle \underbrace{\begin{bmatrix}
    1 & 0 & 1 & 0 \\
    \\
    0 & \lambda_1 & 0 & \lambda_2 \\
    \\
    \lambda_1^2 \cosh{(\lambda_1 L)} & \lambda_1^2\sinh{(\lambda_1 L)} & 
 -\lambda_2^2~\cos{(\lambda_2 L)} & -\lambda_2^2 \sin{(\lambda_2 L)} \\
    \\
    \lambda_1^3 \sinh{(\lambda_1 L)} E_k I  & \lambda_1^3\cosh{(\lambda_1 L)} E_k I & \lambda_2^3 \sin{(\lambda_2 L)} E_k I & -\lambda_2^3 \cos{(\lambda_2 L)} E_k I \\
    + \beta^4 m_{tip} \cosh{(\lambda_1 L)}  & + \beta^4 m_{tip} \sinh{(\lambda_1 L)} &  + \beta^4 m_{tip} \cos{(\lambda_2 L)} & + \beta^4 m_{tip} \sin{(\lambda_2 L)}
    \end{bmatrix}}_{\mathcal{A}} \begin{bmatrix}
    \\
    B_1 \\
    \\
    B_2 \\
    \\
    B_3 \\
    \\
    B_4 \\
    \\
    \end{bmatrix}=\begin{bmatrix}
    \\
    0 \\
    \\
    0 \\
    \\
    0 \\
    \\
    0\\
    \\
    \end{bmatrix}\label{eqn24}
\end{align}
\end{strip}

\noindent
In order to have existence of non-trivial solution, $Det [\mathcal{A}]$ needs to be zero i.e. 
\vspace{-0.1in}
\begin{equation}
    \begin{split}
        D~ & \equiv~ \textit{Det}[\mathcal{A}] = \displaystyle \lambda_1 \lambda_2 \Bigg( \lambda_1^4 + \lambda_2^4 \\
        & + \lambda_1 \lambda_2 \Big( 2 \lambda_1 \lambda_2 \cos{(\lambda_2 L)} \cosh{(\lambda_1 L)}  \\
        & + (\lambda_2^2 - \lambda_1^2) \sin{(\lambda_2 L)} \sinh{(\lambda_1 L)}  \Big) \Bigg) E_k I \\
        & ~-~(\lambda_1^2 + \lambda_2^2) \beta^4 m_{tip} \Big( \lambda_1 \cosh{(\lambda_1 L)}\sin{(\lambda_2 L)} \\
        & - \lambda_2 \sinh{(\lambda_1 L)}\cos{(\lambda_2 L)}\Big)= 0.
    \end{split} \label{eqn25}
\end{equation}

\noindent
After rearrangement, Eq.(\ref{eqn23a}-\ref{eqn23b}) leads to $B_3~=~-B_1$ and $B_4~=~-\displaystyle \frac{\lambda_1}{\lambda_2} B_2$. By substituting  these relationships in Eq.(\ref{eqn23c}), the following equation is derived

\begin{equation}
    \begin{split}
        & B_2 ~=~-\displaystyle \frac{\left(~\lambda_2^2~\cos{(\lambda_2 L)}  
 + \lambda_1^2~\cosh{(\lambda_1 L)} ~\right)}{\left(~ \lambda_1^2~\sinh{(\lambda_1 L)}  + \lambda_1 \lambda_2~\sin{(\lambda_2 L)}~\right)} B_1
    \end{split}
\end{equation}

\noindent
By substituting the solution for $B_3,B_4$ and $B_2$ in Eq.(\ref{eqn22}), $\phi (x)$ can be normalized in terms of $B_1$ as follows:
 
 \begin{equation}
 \begin{split}
    \phi(x)~=~ &  \cosh{(\lambda_1 x)}~-~\cos{(\lambda_2 x)}~ \\
    & -\sigma~\left(\sinh{(\lambda_1 x})~-~\displaystyle \frac{\lambda_1}{\lambda_2}\sin{(\lambda_2x)}~\right) \\
    \\
    \text{where,} ~~ &  \sigma = \frac{\left( \lambda_2^2 \cos{(\lambda_2 L)}  
 + \lambda_1^2 \cosh{(\lambda_1 L)} \right)}{\left(\lambda_1^2 \sinh{(\lambda_1 L)} + \lambda_1 \lambda_2 \sin{(\lambda_2 L)} \right)}
     \end{split}
 \end{equation}

\noindent
If $P = 0 $, the governing PDE in Eq.(\ref{eqn9}) signifies free vibration of an Euler-Bernoulli beam.

\subsection{Formulation of eigenvalue problem}

For different values for $P$ and for all considered mode of vibration, Eq.(\ref{eqn19}) can be rewritten as 
        
        \begin{equation}
        \begin{split}
            & [B] ~\dot{z} (t)~+~[E]~ z (t)~=~0 \\
            \\
            & [B]~=~\begin{bmatrix}
                     ~0~ & ~1~ \\
                     ~1~ & ~0~
             \end{bmatrix}~, \quad [E] ~=~\begin{bmatrix}
                     ~-1~ & ~0~ \\
                     ~0~ & ~\omega^2~
             \end{bmatrix} , \\
             \\
             & \quad \text{and} \quad z (t) ~=~\begin{bmatrix}
                     \dot{q} (t) \\
                     q (t) 
             \end{bmatrix} \label{eqn32}
        \end{split}
        \end{equation}
\noindent        
The general form of solution for $z (\hat{t})$ in Eq.(\ref{eqn32}) is 

\begin{equation}
    z (t)~=~\mathcal{C}~e^{\Lambda t}~=~\mathcal{C}~e^{i \Omega t}.
\end{equation}

\noindent
By substituting the general solution of $z(t)$ in Eq.(\ref{eqn32}) the following algebraic relationship is found

\begin{equation}
\begin{split}
    & [B] ~\Lambda~\mathcal{C}~e^{\Lambda t}~+~[E]~ \mathcal{C}~e^{\Lambda t}~=~0 \\
    & [B]^{-1}~[B] ~\Lambda~\mathcal{C}~e^{\Lambda t}~+~[B]^{-1}~[E]~ \mathcal{C}~e^{\Lambda t}~=~0 ~~ \\
    \\
    & \text{(pre-multiplying LHS with $[B]^{-1}$ )} \\
    \\
    & \left(\Lambda~[I]~+~[B]^{-1}~[E]~\right) \mathcal{C}~e^{\Lambda t}~=~0. 
\end{split}\label{eqn33}
\end{equation}

Eq.(\ref{eqn33}) is a general eigenvalue problem, which needs to be solved at every instance with change of the parameter $P$.

\section{Discussion on Instability} \label{instability}

If the general form of the system described in Eq.(\ref{eqn33}) is rewritten as:

\begin{equation}
\begin{split}
    &   \displaystyle  W (x,t)~=~\sum_{i=1}^N~\phi_i (x) q (t)~=~\sum_{i=1}^N~A_i \phi_i (x) e^{i \omega_i t} \\
    & \text{where} \quad \omega_i \in \mathbb{C}, \quad \text{as in} \quad \omega_i = \pm ( \omega_i^R + i~\omega_i^I ) \\
    \\
    & \displaystyle  W (x,t)~=~\sum_{i=1}^N A_i~e^{\mp \omega_i^I t}~\phi_i (x) \sin{(\omega_i^R t + \theta_i)}
\end{split}
\end{equation}

\begin{enumerate}
    \item When $\omega_i = 0$, $~W (x,t)~=~\sum_{i=1}^N~\phi_i (x)~$. Hence, it is evident that static instability or buckling is only possible when $\omega_i~=~0$.

    \item When $\omega_i$ is repeated,

    \begin{equation}
        \begin{split}
              \displaystyle  W (x,t) & =  \underbrace{C_m e^{-\omega_m^I t} \phi_m (x) \sin{(\omega_m^R t + \theta_m)}}_{I}  \\
            & +  \underbrace{D_ m t~e^{-\omega_m^I t} \phi_m (x) \sin{(\omega_m^R t + \theta_m)}}_{II}\\
            &  + \underbrace{\sum_{i=1, i \neq m}^N A_i e^{-\omega_i^I t} \phi_i (x) \sin{(\omega_i^R t + \theta_i)}}_{III}
        \end{split}
    \end{equation}

Term II is always unstable. Moreover if $\omega_m^R \neq 0 $, the instability is characterized as flutter instability. When $\omega_m^R = 0 $ and $\omega_m^I < 0 $ the instability is characterized as divergence instability.

\end{enumerate}

Similarly, the dynamic instabilities can be characterized from the eigenvalue problem described in Eq.(\ref{eqn33}) as well. $z(t)$ has the form:

\begin{equation}
    \begin{split}
        z (t)~=~\mathcal{C}~e^{-\Im (\Omega) t}~\sin{(\Re (\Omega) t~+~\theta)}
    \end{split}
\end{equation}
\noindent
 Flutter happens when $\Re (\Omega) \neq 0 $ and  $\Im (\Omega) < 0 $. Divergence happens when $\Re (\Omega) = 0 $ and  $\Im (\Omega) < 0 $.

\section{Numerical results}

\begin{figure}[htp]
\centering
{\includegraphics[width=0.4825\textwidth]{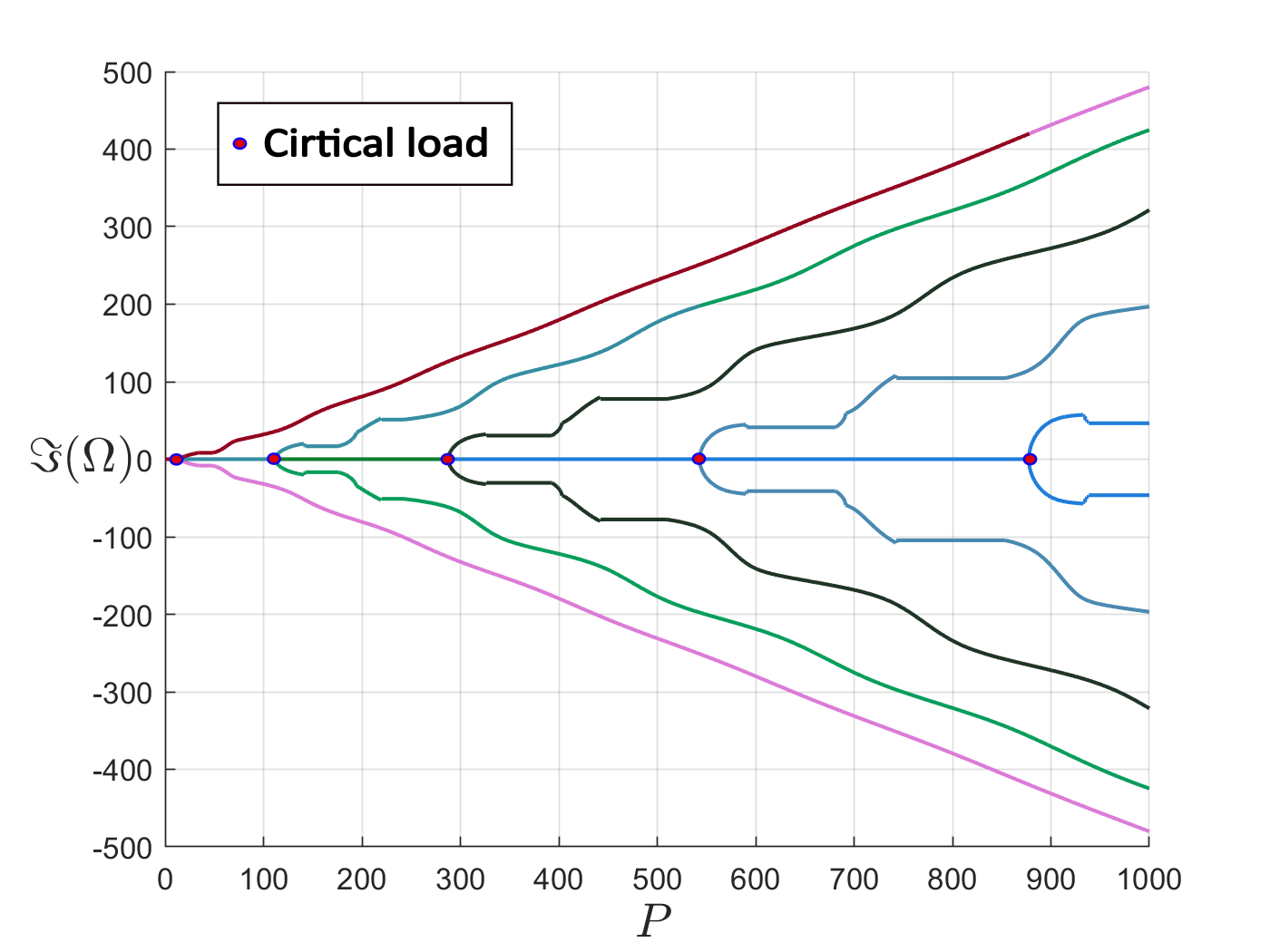}}%
\caption{$P$ vs. $\Im (\Omega)$ plot for $0 \leq P \leq 1000 $. }
\label{figure 3}
\end{figure}

\begin{figure}[htp]
\centering
{\includegraphics[width=0.4825\textwidth]{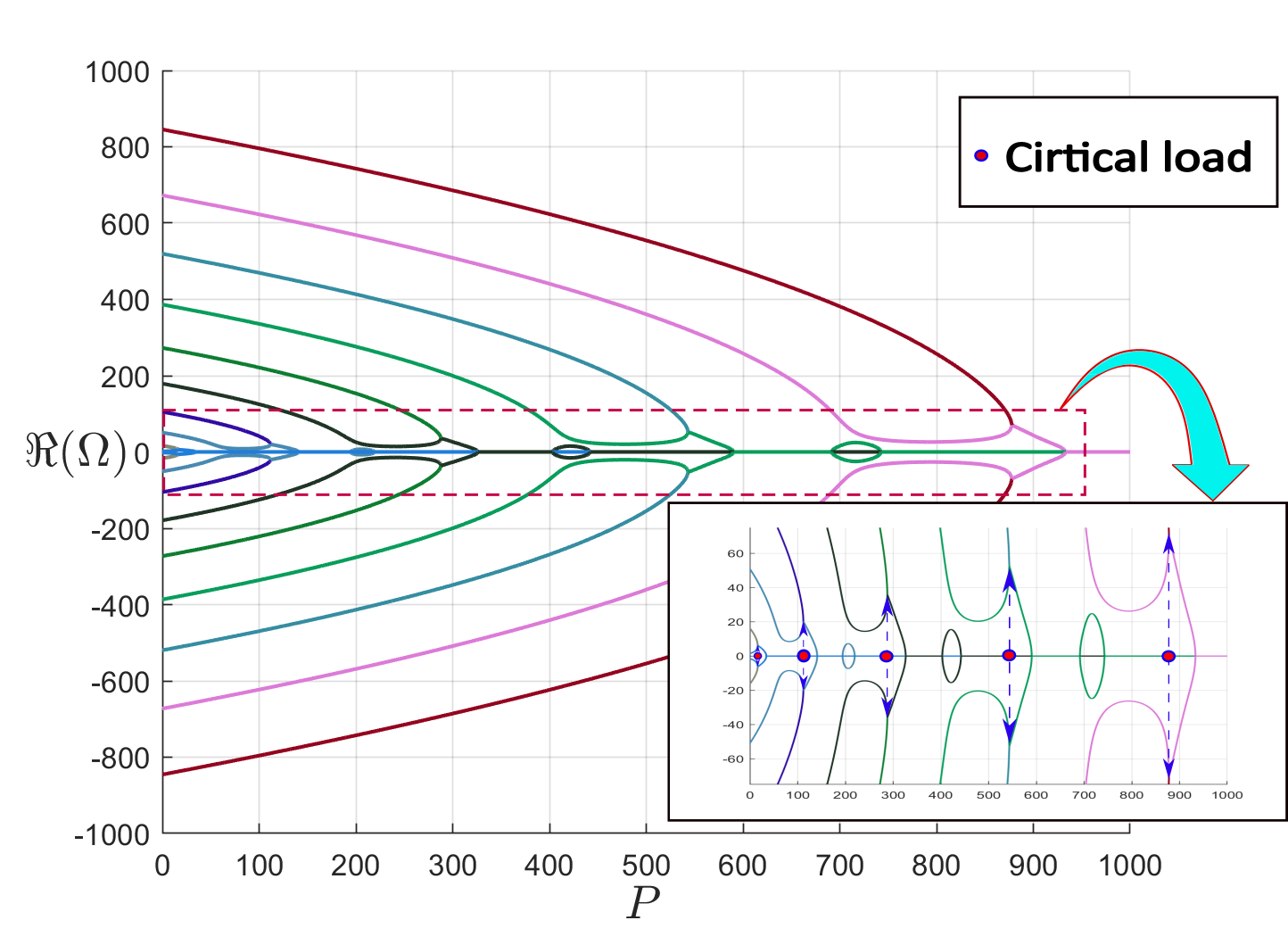}}%
\caption{$P$ vs. $\Re(\Omega)$ plot for $0 \leq P \leq 1000 $. }
\label{figure 4}
\end{figure}

The temporal eigenvalues, $\Omega$ are evaluated for the eigenvalue problem described in Eq.(\ref{eqn33}) for different values for $P$ ($0 \leq P \leq 1000$). In this particular example, only the first 10 modes are considered. Hence Eq.(\ref{eqn33}) yields 10 pairs of complex eigenvalues and their respective complex conjugates. Fig.(\Ref{figure 3}-\ref{figure 4}) show the evolution of $\Im (\Omega)$ and $\Re (\Omega)$ with respect to different values for $P$. As discussed in the previous section, both types of instabilities i.e. flutter and divergence are observed at different interval for $P$. An important observation can be made from Fig.(\ref{figure 4}) that $\Re (\Omega)$ associated with two consecutive modes coincide with each other at the same values for $P$ as bifurcations begin in Fig.(\ref{figure 3}). 

\begin{figure}[htp]
\centering
{\includegraphics[width=0.4825\textwidth]{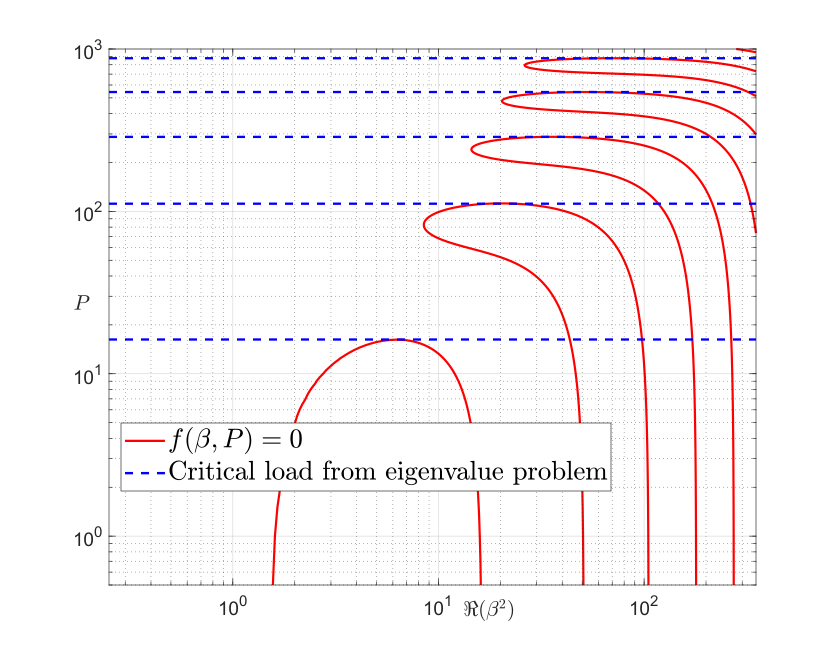}}%
\caption{Natural frequency $\Re (\beta^2) = \omega^R$ vs. $P$ plot.}
\label{figure 5}
\end{figure}

For different values for $\omega^R$, $P$ is evaluated from the nonlinear algebraic equation presented in Eq.(\ref{eqn25}). Results are presented in Fig.(\ref{figure 5}). Results presented in Fig.(\ref{figure 5}) also corroborate the observation from Fig.(\ref{figure 4}). At $d P/d \omega^R = 0$ i.e. when $\omega^R$ is repeated bifurcation curves begin to develop in Fig.(\ref{figure 3}). In Fig.(\ref{figure 5}) $\omega^R$ is always nonzero i.e. only dynamic instability can occur in the system.

\addtolength{\textheight}{-12cm}   





\section{Conclusions}

The stability analysis of a Euler-Bernoulli beam with a tip mass and a follower force has been analyzed. The exact eigenfunctions which satisfy the natural and geometric boundary conditions are used to arrive at the characteristic equations to study the locus of the temporal eigenvalues of the beam dynamics. Two approaches are used to validate the initiation of flutter instability which corresponds to the convergence of two distinct eigen-frequencies. These results are currently being validated with an experimental testbed.

\bibliographystyle{./IEEEtran}
\bibliography{IEEEabrv, EBfollowerTipMass}

\end{document}